\documentclass[12pt]{article}
\usepackage{amsmath,amssymb,graphicx,mathrsfs,hyperref,slashed,setspace}

\usepackage{float}
\usepackage{latexsym}
\usepackage{bm}
\usepackage{blindtext}
\hypersetup{
	colorlinks=true,
	linkcolor=blue,
	filecolor=black,
	urlcolor=black,
	citecolor=blue,
}

\newcommand{\hide}[1]{}

\newcommand{\be}{\begin{equation}}
\newcommand{\ee}{\end{equation}}
\newcommand{\bea}{\begin{eqnarray}}
\newcommand{\eea}{\end{eqnarray}}

\newcommand{\cK}{{\cal K}}
\newcommand{\cH}{{\cal H}}
\newcommand{\cP}{{\cal P}}
\def\({\left(}
\def\){\right)}

\begin{document}
%%%%%%%%
\title{\vspace{-1.8in}
%\vspace{3mm}
%\vspace{0.3cm}
{Kerr Black hole mimickers \\ sourced by a string fluid}}
\author{\large Ram Brustein${}^{(1)}$,  A.J.M. Medved${}^{(2,3)}$, Tamar Simhon${}^{(1)}$
\\
\vspace{-.5in} \hspace{-1.5in} \vbox{
\begin{flushleft}
 $^{\textrm{\normalsize
(1)\ Department of Physics, Ben-Gurion University,
   Beer-Sheva 84105, Israel}}$
$^{\textrm{\normalsize (2)\ Department of Physics \& Electronics, Rhodes University,
 Grahamstown 6140, South Africa}}$
$^{\textrm{\normalsize (3)\ National Institute for Theoretical Physics (NITheP), Western Cape 7602,
South Africa}}$
\\ \small \hspace{0.57in}
   ramyb@bgu.ac.il,\  j.medved@ru.ac.za,\ simhot@post.bgu.ac.il
%\ Tom's-Email,
%\ Tamar's-Email
\end{flushleft}
}}
\date{}
\maketitle
%%%

%%%%%%%%%%%%%%
%\renewcommand{\baselinestretch}{1.15}
%%%%%%%%%%%%%%%%%
%%%%%%%%%%%%%%%%%%
\begin{abstract}
%%%%%%%%%%%%%%%%%%%
%\abstract{

We present  rotating solutions of Einstein's gravity coupled to an effective Born-Infeld theory that describes the end of open-string tachyon condensation after the decay of an unstable $D$-brane or a brane-antibrane system.  The geometry of these  solutions is that of  the rotating frozen star. The solutions are stationary, non-singular and ultracompact, and their exterior geometry is identical, for all practical purposes, to that of the Kerr solution.  The Born-Infeld matter consists of radial electric-flux tubes that emanate from, or end at, the ellipsoidal core of the star.   Each end of the flux tubes carries an electric charge, so that the electric field can be viewed as being sourced by an ellipsoidal charge distribution of positive and negative charges near the center of the star. Meanwhile,  the star's  outer layer is equal and oppositely charged, resulting in a vanishing electric field in the external spacetime.  We show that these rotating solutions are ultrastable against radial perturbations, just like their static frozen star counterparts. They are also effectively immune to ergoregion   instabilities, as we discuss.

%EXPLAIN VALUE ???????????????????

%%%%%%%%%%%%%%%%%
\end{abstract}
%%%%%%%%%%%%%%%%%%%%%%%%
%\maketitle
\newpage
%%%%%%%%%%%%%%%%%%%
\setstretch{1.5}
%%%%%%%%%%%%%%

\renewcommand{\baselinestretch}{1.5}
\section{Introduction}

The frozen star model \cite{bookdill,BHfollies,popstar,trajectory,fluctuations} describes a type of ultracompact object \cite{carded} that is horizonless and regular but, yet,  able to mimic a black hole (BH) when viewed by an external observer. Initially, we used it to describe a spherically symmetric and static solution of Einstein's equations and thus a Schwarzschild mimicker. In addition to mimicking the classical aspects of a Schwarzschild BH, the frozen star model was  also able to reproduce the thermodynamic properties of an equal-mass Schwarzschild BH  \cite{U4Euclidean}. The frozen star is sourced, effectively, by a
fluid of maximally negative radial pressure and vanishing transverse pressure throughout the bulk of the interior.~\footnote{The ``bulk of the interior'' is shorthand for everywhere besides a small regularization zone near the center and a  thin layer adjacent to  the outermost surface. Here, ``small'' and ``thin'' means scales on the order of the string length or the Planck length, which are much smaller than the radius of the star.}
This is, of course, rather exotic matter, but its motivation
follows from an earlier proposal for describing the interior of a BH mimicker  --- the polymer model \cite{strungout,emerge}  --- in which case, the object consists of a collection of long, closed, interacting, fundamental strings that are excited to extremely hot temperatures. The idea is that the frozen star model describes the effective  classical geometry and semiclassical aspects that an external observer would attribute to such a BH mimicker, even if she is ignorant about its microscopic description.

The polymer model is premised on the notion that a BH mimicker needs to contain matter that is
in a strongly non-classical state so as to account for the maximally entropic state of the object. A consequence of such highly quantum matter is that the interior does not have a classical geometrical description.  The frozen star geometry is then meant to capture the main features of the polymer interior but from a classical and semiclassical perspective. This leads to some unorthodox  aspects of the frozen star solution, including the highly anisotropic state of matter  as described above, as well as the property that every radial surface in the bulk of the interior is
a surface of very large (but finite) redshift, same as for the would-be horizon. As unconventional
as these properties may be, they do conspire to circumvent singularity theorems \cite{PenHawk1,PenHawk2},
constraints on the compactness of matter \cite{Buchdahl,chand1,chand2,bondi} and issues
with energy conservation when accounting for the effects of evaporation \cite{frolov,visser}.

Importantly, another microscopic description of
the polymer model has recently been identified and presented in \cite{fluxUstat}. The  same frozen star geometry was shown  to be  sourced by a fluid of cold, open strings resulting from the decay of an unstable $D$-brane or a brane--antibrane system at the end of open-string tachyon condensation. The effective Lagrangian for this process  was originally  formulated and developed  by Sen  \cite{Sen,Sen2,Sen3,Sen4,Sen5} (and see \cite{SenReview} for a review with more references).
 Later on, Sen's Lagrangian was recast by
 Gibbons, Hori and Yi (GHY) \cite{GHY} (also see \cite{Yi}) into  a  specific Born--Infeld form that is describing a  fluid of rigid electric-flux tubes (equivalently, open strings).  A similar form of Lagrangian with a two-form field had already appeared in earlier literature in the context of describing the physics of flux tubes and strings: firstly  in \cite{NO} and then later with regard to  the cloud-of-strings model \cite{Letel}, the string-dust model \cite{Stach}  and  ``hedgehog'' compactifications    \cite{hedge1,hedge2}.

When gravity is neglected, the Born--Infeld theory of $Dp$-branes leads to  BIons, which are  spherically symmetric, static and  solitonic solutions of finite energy \cite{GibbonsBIons,GibbonsRev}. These BIons
can be viewed as flux tubes that extend from a central point-like source all the way  out to infinity. It follows from this point of view that  frozen stars  can  be viewed as  gravitationally back-reacted BIons or alternatively, as an example of the   gravitational confinement of open strings.
What we have found in \cite{fluxUstat} is that the outermost surface of the star  contains an equal and opposite
charge that cancels  out the point-like source in the center; meaning that the flux tubes terminate
at the outer surface, so that both the energy and spatial extent of these BIons are finite. In spite of the attractive force between the  opposite charges, stability is
maintained because of an effective force that is associated with the  constraint
 of constant mass.

Unlike the polymer model, the Born--Infeld framework provides us with a well-defined matter Lagrangian to couple to that of Einstein's theory. This completes the Einstein equations and allows us to describe the dynamics of the frozen star in a generic state  that is not necessarily static nor spherically symmetric. Therefore, we can now claim that the frozen star is a consistent and complete model which describes an ultracompact object whose equilibrium state is, for all practical purposes, indistinguishable from that of a Schwarzschild BH for an external observer.

The Born--Infeld framework should  eventually enable us to study extensions  of our model to situations  in which the  departures from equilibrium physics are macroscopically large, as would be the case for an analysis  on  the dynamics of  astrophysical BH mergers.   As discussed in \cite{collision,ridethewave}, this is just what is needed  if one is to observationally  distinguish frozen stars from Schwarzschild BHs, as well as  from  other BH mimickers.

In this paper, we have a more modest extension in mind, namely, to the case of a rotating (but stationary) frozen star solution; that is, to a Kerr BH mimicker. This can be viewed as a critical first step  in a program for extending   the frozen  star solution
beyond  the static, spherically symmetric case.  The inclusion of rotation is a vitally important step in our program as it allows us to  make contact with observational physics.
We have, in fact,  already managed to incorporate rotation into the original frozen star
framework \cite{notstevekerr}.~\footnote{For other attempts at incorporating rotation into
BH mimickers of various types, see \cite{X0,X00,X000,X1,X2,X3,X4,X5,X6,X7,X8,X9,X10,X11,X12,X13,X14,X15,X16,X17,X18,Y1,Y2}
for a partial list. One can also consult \cite{W0,Y0} for reviews.}
This task  was accomplished by extending the solution to the case of non-vanishing spin parameter  $a$, which is the analogue of the standard Kerr spin parameter,
in a way that preserves the basic properties of the frozen star while maintaining the symmetries of the Kerr solution such as the constancy of
the Carter constant \cite{carter} (which is not a constant {\em a priori}).
There was still some degree of arbitrariness
in the procedure, but we negated  this  by verifying that the volume integrals of appropriate elements
of the stress tensor are able to reproduce   the correct mass and angular momentum of a Kerr BH with the same parametrizations.

The main purpose of the current paper is to show explicitly that the Born--Infeld framework includes
the rotating solution as discussed above (see Sec.~3.2).  In addition, we will establish that the rotating frozen star
is ultrastable (Sec.~4.1), just like its static counterpart, and is essentially immune  to  the type of ergoregion instabilities that typically afflict rotating, horizonless, compact objects (Sec.~4.2). We will also discuss
how the internal geometry is matched to the external
Kerr geometry at the star's  outer surface (Sec.~5.1) and how it is regularized at the star's center (Sec.~5.2). Some details about the matching  procedure will be deferred to an appendix. Section~2 provides a review of \cite{fluxUstat}, whereas
Section~3.1 is a review of \cite{notstevekerr}.

\section{Lagrangian and equations of motion}

In this section, we recall formalism and  results from  \cite{fluxUstat} for the sake of self-completeness and to establish our notations. Many of the expressions below first appeared
in \cite{GHY}. Our conventions are that an index of  0 denotes  time, one of $a$,$b$,$\cdots$ denotes an arbitrary spacetime dimension   and  one of $i$,$j$,$\cdots$ denotes an arbitrary  spatial
dimension. Three large spatial dimensions are assumed
for concreteness.
%and, for spherical coordinates, $\;(0,1,2,3)=(t,r,\theta,\phi)\;$.

Our model describes Einstein's gravity coupled to a Born--Infeld-like theory for the antisymmetric tensor ${\cK}_{ab}$. The action is, up to possible surface terms (including
a constant-mass constraint),
\be
S_{GBI}\;=\;\int d^4 x \left\{\frac{1}{16 \pi G} \sqrt{-g}R +\frac{1}{2\pi \alpha'}\sqrt{-\frac{1}{2}{\cal K}^{ab}{\cal K}_{ab}}+\lambda_1 \varepsilon^{abcd}{\cal K}_{ab}{\cal K}_{cd} +  \sqrt{-g} J_a A^a \right\} \;.
\label{GBI}
\ee
The term with the Lagrange multiplier $\lambda_1$ and the source $J_a$ will be discussed later.
To arrive at this action, we followed the analysis of GHY \cite{GHY} (also, \cite{Yi}),
whose own starting  point was Sen's effective action for the final state of the decay of unstable  $D$-branes at the end of  tachyon condensation and near the  minimum  of the tachyon potential where it vanishes \cite{Sen,Sen2,Sen3,Sen4,Sen5}.

The connection with flux tubes or open strings follows from the observation in \cite{GHY} that ${\cal K}^{ab}$ can be identified as a surface-forming bivector, which can always be regarded as the cross-sectional slice  of the world sheet of an open string.

\subsection{Lagrangian}

Sen's effective Lagrangian~\footnote{Here and elsewhere, we use the terms Lagrangian, Hamiltonian, {\em etc}. instead of Lagrangian density, Hamiltonian density, {\em etc}.  to reduce clutter.} takes the following form:
\be
{\cal L}\;=\;- V(T)\sqrt{-Det\left(g+ 2\pi \alpha'{\cal F}\right)}\;+\;
\sqrt{-g}
A^a J_a\;,
\label{LofT}
\ee
where $V(T)$ is  both the $D$-brane  tension and the tachyon potential,
 $g_{ab}$ is the background metric, $2\pi \alpha'$ is the inverse of the fundamental string tension,
$\;{\cal F}_{ab}=\partial_{a}A_{b}-\partial_{b}A_{a}\;$ is the field-strength  tensor for  the gauge field $A_a$ and $J_a$, a 4-vector current density, is the source.

Although the Lagrangian~(\ref{LofT}) vanishes when $V(T)$ vanishes, the
corresponding Hamiltonian
\be
{\cal H}\;=\;\frac{\delta {\cal L}}{\delta  (\partial_0 A_i)}-{\cal L}\;=\;\frac{1}{2\pi\alpha'}\sqrt{D^i D_i +  {\cal P}^i{\cal P}_{i}}\;,
\ee
remains well defined. Here, ${\cal P}_i$ is the conserved momentum (see below),
$\;E_i={\cal F}_{0i}\;$  is the electric field and
$D^i$ is the electric displacement, which is  defined as the canonical conjugate of the gauge field and thus  as $\;D^i=\frac{\delta {\cal L}}{\delta (\partial_0  A_i)}\;$.
As is commonly the case in  Born--Infeld theories, it is $D^i$ and not $E^i$ which satisfies the Gauss'-law constraint ($\rho_e$ is the volume charge density),
\be
\nabla_i D^i\;=\;J_0\;=\;\rho_e\;.
\label{sourceD}
\ee
Also, in the flux-tube picture, it is $D_i$ that determines the direction of the electric flux.

The conserved momentum ({\em i.e.}, the analogue of ${\cal H}$ for spatial translations)
is expressible as
\be
\;\frac{1}{2\pi\alpha'}{\cal P}_i\;=\;-{\cal F}_{ij} D^{j}\;=\;\left(\vec{D}\times\vec{B}\right)_i\;,
\label{cPdef}
\ee
where $B_i$ is the magnetic induction and defined by
\be
 \;B^i\;=\;\frac{1}{2}\varepsilon^{ijk} {\cal F}_{jk}\;.
\ee

Along with  the Gauss'-law constraint~(\ref{sourceD}), the fields are required to satisfy
Ampere's law $\;\nabla_i {\cal F}^{i}_{\;\;j}- \nabla_0{E}_j= J_j\;$
and the Bianchi identities, which include Faraday's law $\;{\vec{\nabla} \times}\;\vec{E}+\partial_0{\vec{B}}=0\;$ and, given that  there are no sources for magnetic monopoles,
\be
\nabla_i B^i\;=\;0\;.
\label{sourceBD}
\ee

To obtain a Lagrangian that can be used near the minimum of the potential, GHY changed variables from the elements
${\cal F}_{ij}$ to their duals $\;K^{ij}=2\frac{\delta {\cal H}} {\delta {\cal F}_{ij}}\;$. The fields ${\cal F}_{ab}$ and, as defined in Eq.~(\ref{KalK}), ${\cal K}_{ab}$
should be regarded as independent variables. This is important when the equations of motion (EOM) for ${\cal K}_{ab}$ are derived.

A new  Lagrangian can  be defined by using  a  Legendre transformation of the form
\be
{\cal L}'\;=\; {\cal H}-\frac{1}{2}{\cal F}_{ij}K^{ij}\;=\;\frac{1}{2\pi\alpha'}\sqrt{D^iD_i-\frac{1}{2}K^{ij}K_{ij}}\;,
\label{canon}
\ee
with the latter equality following by way of the  identity
\be
 \frac{1}{2}K^{ij}K_{ij}\;=\;\frac{1}{(2\pi\alpha'{\cal H})^2}{\cal P}^{i}{\cal P}_{i}D^j D_j\;.
 \label{thekey}
\ee

It is useful to introduce an anti-symmetric two-form field ${\cal K}_{ab}$ that acts as an effective field strength,
\bea
{\cal K}_{0i}&=& D_i
\;, \cr
{\cal K}_{ij}&=& K_{ij}\;,
\label{KalK}
\eea
so that the  new Lagrangian~(\ref{canon}) can be written as one that is manifestly
of the Born--Infeld form
\be
{\cal L}'\;=\; \frac{1}{2\pi \alpha'}\sqrt{-\frac{1}{2}{\cal K}^{ab}{\cal K}_{ab}}\;.
\label{newL}
\ee
Here,  it has been assumed that the electric-field term dominates over the magnetic one and the negative sign is a consequence of $\;g_{tt}<0\;$ appearing in the contraction when $\;a=0\;$ and $\;b=i\;$ (and {\em vice versa}). In the absence of magnetic sources,
the  new Lagrangian is equal to the original Hamiltonian and simplifies to
$\;{\cal L}' = {\cal H} = E_iD^i \;$.

There is a technical issue: A canonical analysis of ${\cal L}'$ does not lead back to the same Hamiltonian ${\cal H}$.
However, as explained  in \cite{GHY}, this
 can be  corrected by using a Lagrange multiplier (here, denoted by $\lambda_1$)   to  impose the constraint
 $\;{\cal K}\wedge {\cal K}=0\;$, which follows directly from Eq.~(\ref{KalK}).
This accounts for the Lagrange-multiplier term in Eq.~(\ref{GBI}).

Another  subtlety concerns the source term.  The source term in Eq.~(\Ref{LofT}) is unaffected by the Legendre transformation; however, $A_a$ is not the gauge field for the field strength ${\cal K}_{ab}$.  But, since  ${\cal F}$ and $\cK$ are to be regarded as independent, one can  vary the  Lagrangian ${\cal L}'$ with respect to either field. The variation with respect to
$A^{a}$ leads to the actual  Gauss'-law constraint as in Eq.~(\ref{sourceD}).
This is quite obvious in the case of no magnetic sources because, then,   $\;{\cal L}' = E_iD^i \;$.

\subsection{Equations of motion}

The EOM for $\cK$ are particularly transparent when expressed in terms of the original field-strength tensor, as these are the same as  the original  Bianchi identities,
$\;
d{\cal F}=0\;
%\label{BIEOMa}
$.
Meanwhile, the  Bianchi identities for the new field-strength tensor
$\;
 d{\cal K}=0\;
 %\label{BIEOMb}
 $
happen to be the same as   the EOM for the original Lagrangian.

The gravitational EOM are, as always,
\be
\frac{1}{8\pi G}G^{a}_{\;\;b}\;=\; T^a_{\;\; b}\;.
\label{Ein}
\ee
For the geometry of interest,  the sources are restricted  to the core and the outer surface of the star.  In this case,  as justified in \cite{Letel}, the stress tensor works out as follows:
 \bea
 T_{ab} \;=\; 2\frac{\delta {\cal L}'}{\delta g^{ab}}
        \;=\; \frac{1}{2\pi\alpha'}\frac{{\cal K}_{a}^{\;\;c}{\cal K}_{bc}}{\sqrt{-\frac{1}{2}{\cal K}^{ab}{\cal K}_{ab}}} \;.
 \label{TBI}
 \eea
Alternatively, in terms of the effective electric field and the magnetic field via the conserved momentum, the stress tensor is given by
\bea
T_{00} &=&  \frac{1}{2\pi\alpha'}\frac{D^i D_i}{\sqrt{D^i D_i -\frac{1}{2}K^{ij}K_{ij}}}\;=\; {\cal H}\;, \label{Too} \\
T_{0i}  &=&    \frac{1}{2\pi\alpha'}\frac{D^j K_{ij}}{\sqrt{D^i D_i -\frac{1}{2}K^{ij}K_{ij}}}\;=\; -\frac{1}{2\pi\alpha'}{\cal P}_i\;,
\label{Toi}\\
T_{ij} &=&  \frac{1}{2\pi\alpha'}\frac{-D_i D_j+K_{i}^{\;\;k}K_{jk}}{\sqrt{D^i D_i -\frac{1}{2}K^{ij}K_{ij}}}\;=\;
\frac{1}{(2\pi\alpha')^2}\frac{-D_iD_j+{\cal P}_i{\cal P}_j}{{\cal H}}\label{Tij}\;,
\eea
where the right-most equalities can be obtained by way of Eq.~(\ref{thekey}).

\section{Rotating solutions}

The rotating frozen star has the same exterior geometry as a Kerr BH but the interior
geometry  is much different, as we will now proceed to describe.

\subsection{The metric and Einstein tensor}

Let us begin here by recalling from \cite{notstevekerr} the interior (bulk) metric for  the rotating frozen star in  corotating or zero-angular-momentum-observer (ZAMO) coordinates \cite{MTW,frolovtimes2}. These are obtained from  Boyer--Lindquist coordinates by the transformation $\;t\to \chi\;$ and $\;\phi \to \phi+ \omega\chi\;$, where $\omega =\tfrac{a}{r^2+a^2}$ is the local angular velocity, and  $a$ has the same meaning as the spin parameter for the Kerr solution and a dimensionality of length. This transformation ensures that each radial slice  in the bulk of the interior is a nearly null surface for which redshift is large but finite. This is the same geometry but with  a different radial size  as  that of  the star's outermost surface.   The area of any such  slice  is $\;A(r)= 4\pi R^2=4\pi (r^2+a^2)\;$, with $R$ being the Kerr areal coordinate. The would-be Kerr horizon is located at $\;R_+=\sqrt{r_+^2 +a^2}\;$, with $\;r_+=M(1+\sqrt{1-a^2/M^2})\;$, where
$M$ is the mass of the star.

The interior metric is quite simple,
\be
ds^2_{FS/ZAMO} \;=\; -\varepsilon^2 d\chi^2 + \frac{dr^2}{\varepsilon^2}
+\Sigma d\theta^2 +\frac{R^4}{\Sigma} \sin^2{\theta} d\phi^2 \;,
\label{diagonal2}
\ee
where $\;\varepsilon^2\ll 1\;$ is a dimensionless parameter  and
$\;\Sigma=r^2 + a^2 \cos^2 \theta\;$.  It should be
noted that $g_{\phi\phi}$ differs from that of  Boyer--Lindquist Kerr coordinates.  This is because,
in the spirit of the original frozen star model, the mass $M$  of the star
is redefined by a scaling relation:~\footnote{In spite
of the rescaling, $M$ will still refer to the total mass of the star unless
an explicit functional dependence is specified.} $\;M=\frac{R_+^2(1-\varepsilon^2)}{2r_+}
\to M(r)=\frac{R^2(1-\varepsilon^2)}{2r} \;$. Note that $M$, like $a$, has a dimensionality
of length and the factor of  $1-\varepsilon^2$ accounts for the fact that the outermost surface
extends slightly past the location of the would-be Kerr horizon (we are neglecting order-$\varepsilon^4$
corrections).

On the basis of the antecedent polymer model,
we expect that the small parameter $\varepsilon^2$ --- which has been called $\epsilon$ or $\varepsilon$ in some earlier papers   ---   scales as the ratio of the Planck length to the star's radius. This ratio is $10^{-38}$ for a solar-mass ultracompact object, so that $\varepsilon^2$  is extremely small.

The metric~(\ref{diagonal2}) is valid for all values of $a$  such that $\;\frac{a}{M}\le 1\;$
and limits to that  of the static frozen star as $\;a\to 0\;$.
That the  limiting value of $a$ is equal to  its limiting value for the  Kerr solution
can be understood as follows:
For our solution, $\;a\left(r\lesssim r_+\right)\leq a\left(r\gtrsim r+\right)\;$; otherwise, the star will shed  mass until a stationary state is attained. It then follows that $a_{interior} \leq a_{exterior}\leq M\;$.

ZAMO coordinates are often convenient to use but hide the fact that the star is rotating. This
aspect, however, becomes  apparent with the restoration of the original $t$ and  $\phi$ coordinates. This leads to the frozen star metric  in its analogue of
Boyer--Lindquist coordinates,
\bea
ds^2_{FS}&=&\left(\frac{a^2\sin ^2\theta}{\Sigma}-\varepsilon ^2 \right)dt^2 \;-\;2a\sin ^2\theta \frac{ R^2 }{\Sigma} dtd\phi
\;+\; \frac{1}{\varepsilon ^2} dr^2
\nonumber \\
\;&+&\; \Sigma d\theta^2\;+\;\frac{R^4 }{\Sigma}\sin ^2\theta d\phi^2\;.
\label{offdiagonal1}
\eea

It is useful to compare the frozen star metric in Eq.~(\ref{diagonal2}) to the  near-horizon limit of the  Kerr metric in ZAMO coordinates. To leading order in $\;\Delta=r^2- 2Mr+a^2\;$,
the latter  metric can be expressed as
\be
ds^2_{Kerr/ZAMO} \;=\; -\frac{\Delta}{\Sigma} d\chi^2 + \frac{\Sigma}{\Delta} dr^2
+\Sigma d\theta^2 +\frac{R^4}{\Sigma} \sin^2{\theta} d\phi^2\;.
\label{diagonal3}
\ee
Since $\Delta$ vanishes at the horizon where $\;r=r_+\;$,
it is clear that the frozen star model  essentially trades $\frac{\Delta}{\Sigma}$ for $\varepsilon^2$ and then extends this near-horizon form  throughout the bulk of the interior.
It is also clear from Eqs.~(\ref{diagonal2}) and~(\ref{diagonal3}) that the metric is
continuous at the outer surface; however, its derivatives are not. We correct this later on  by introducing a thin transitional layer which interpolates smoothly between the interior and exterior geometries.

 The frozen star geometry  is regular throughout the interior except for a removable, mild singularity at the center of the star. As discussed later on, this can be regularized
 by smoothing the metric in a small region near the core of the star. This regularization, as well as the smoothing procedure at the outer surface,  does not change any of our results in a significant way. It only changes physical quantities,  such as the mass, angular momentum  and charge, by  perturbatively small amounts.

The Einstein tensor $\;G_{\mu\nu}={\cal R}_{\mu\nu}-\frac{1}{2} g_{\mu\nu}{\cal R} \;$ has the following diagonal components to leading order in $\varepsilon^2$:
\be
G^{t}_{\;\;t}\;=\;G^{r}_{\;\;r}\;=\; -\frac{R^2\left(r^2-3a^2\cos^2{\theta}\right)}{\left(r^2+a^2\cos^2{\theta}\right)^3}\;,
\label{GttGrr}
\ee
\be
G^{\theta}_{\;\;\theta}\;\;,\; \;G^{\phi}_{\;\;\phi}\;=\; 0\;.
\label{GththGphph}
\ee
There is also
 a pair of non-vanishing off-diagonal elements, $G_{r\theta}$ and its inverse.
 But these  can be removed by diagonalizing the $r$--$\theta$ sector, leaving the diagonal components  unchanged up to $\varepsilon^2$ corrections.

One can see that $\rho$ is negative in a region around the axis of rotation where $r^2-3 a^2 \cos^2 \theta $ is negative. This does not lead to issues related to causality as discussed at length in \cite{notstevekerr}.

\subsection{Born-Infeld fields and sources}

When $\;a\neq  0\;$, the simplest solution of the EOM is an axially symmetric and apparently  static  configuration. The solution looks like a static solution  because of our choice to use  ZAMO coordinates, but it  ``knows'' about the rotation and so is ``secretly'' stationary.  For this solution,
$\;{\cal K}_{tr}=-{\cal K}_{rt}\;$ and all other elements  of ${\cal K}$  vanish. One then finds that  $\;{\cal K}^{ab}{\cal K}_{ab}=-2 D^r D_r\;$.  Then,
just as in \cite{fluxUstat} for  the   spherically symmetric case,  the only non-vanishing elements of $T_{ab}$ are
\be
-T^t_{\ t} \;= \;T^r_{\ r}\; =  \;\frac{1}{2\pi\alpha'}\sqrt{D^r D_r}\;.
\label{SS}
\ee
As expected, $\;p_r=-\rho\;$ and $\;p_\perp=0\;$, which  corresponds to the so-called  string  fluid  as described in \cite{GHY}.

The Einstein equations~(\ref{Ein}) along with Eqs.~(\ref{GttGrr})  and~(\ref{SS}) can be used to obtain
\be
\frac{1}{8\pi G} \frac{R^2\left(r^2-3a^2\cos^2{\theta}\right)}{\left(r^2+a^2\cos^2{\theta}\right)^3} \;=\;   \frac{1}{2\pi\alpha'}\sqrt{D^r D_r}\;.
\label{singlepeak}
\ee
If we adopt the conventions that $\;D^r=D_r\;$~\footnote{This convention is for convenience but also follows from the relation $\;D^rD_r=\frac{1}{2}K^2_{rt}\;$  because
$\;K^{rt}=-K_{rt}\;$.} and that $\;D_r>0\;$ in the $\;a\to 0\;$ limit,
then the magnitude of the electric displacement $D_r$
can be written as
\be
 D_r \;=\; \frac{\alpha'}{4 G} \frac{R^2\left(r^2-3a^2\cos^2{\theta}\right)}{\left(r^2+a^2\cos^2{\theta}\right)^3}\;,
\label{solD}
\ee
whereas $D_{\theta}$, $D_{\phi}$ are vanishing.~\footnote{It may seem counter-intuitive
that $D_r$ can depend on $\theta$ while $D_\theta$ can be zero. However, the {\em effective}
field-strength tensor ${\cal K}$ does not necessarily  have to be defined  in terms of
an unconstrained gauge field, as the source terms depend only on the original gauge field that
is associated with the field strength ${\cal F}$ \cite{GHY}.}
In the region where $\;D_r<0\;$, the electric flux is  ingoing towards the center, while in regions of positive $D_r$, the flux is radially outgoing. The configuration of the flux is depicted in Fig.~1 for several values of $a$.

The conservation of $T_{ab}$ and the antisymmetry properties of $\cK_{ab}$ imply that, in the bulk of the frozen star,
$\;
\nabla_i D^i =0
\; $,
as discussed in Section III  of \cite{Letel} and as can be  verified  by  a direct calculation.
It follows that there can be no charges located anywhere in the bulk.
This observation supports the  flux-tube picture, which  helps in understanding the  location  of the charges. The end of each  flux tube  carries a charge  that is equal in magnitude and opposite in sign to the charge on the opposite end, so that the total charge of the flux tube vanishes. This charge distribution can be  substantiated by calling on the Gauss'-law constraint~(\ref{sourceD}) and then choosing a narrow, radially directed Gaussian surface that  caps off at  each end of the flux tube.  The rotating frozen star can therefore be viewed as being sourced by a fluid of electric-flux lines. Some of the flux lines are emanating radially from a source in the central core and extending all the way to the surface of the star. Another group of flux lines are emanating from the special surface where $D_r$ vanishes, $\;r^2-3 a^2 \cos^2\theta=0\;$,  and ending at the center.  Yet another group of flux lines emerge from the surface of vanishing $D_r$,  extending  from there to the outer surface. This is depicted in Fig.~1, where the arrows are pointing at the start and end points of the flux lines in the different regions.

According to Eq.~(\ref{solD}), the flux tubes are radial, but the coordinate $r$ is part of an oblate spheroidal coordinate system and not a standard radial coordinate \cite{DrawingKerr}.
The oblate spheroidal coordinates  are related to  standard Cartesian coordinates as follows:
\bea
x &=& R \sin\theta \cos\phi\;,  \cr
y &=& R \sin\theta \sin\phi\;, \\
z &=& r \cos\theta \;.  \nonumber
\eea

The radial direction for  fixed $\theta$ and $\phi$ coordinates is a hyperbola,
\be
\frac{x^2+y^2}{a^2 \sin^2\theta}-\frac{z^2}{a^2 \cos^2\theta}\;=\;1\;,
\ee
while cross sections of surfaces of fixed $R$ are confocal ellipses,
\be
\frac{x^2+y^2}{R^2}+\frac{z^2}{r^2}\;=\;1\;.
\ee
The surface at $\;r=0\;$ is a degenerate ellipsoid whose height (range of the $z$ coordinate) goes to zero with $r$ but whose surface area, $\;A=4\pi a^2\;$,  nevertheless remains finite.

\begin{figure}[t]
\vspace{-.95in}
\begin{flushright}
\includegraphics[height=8cm]{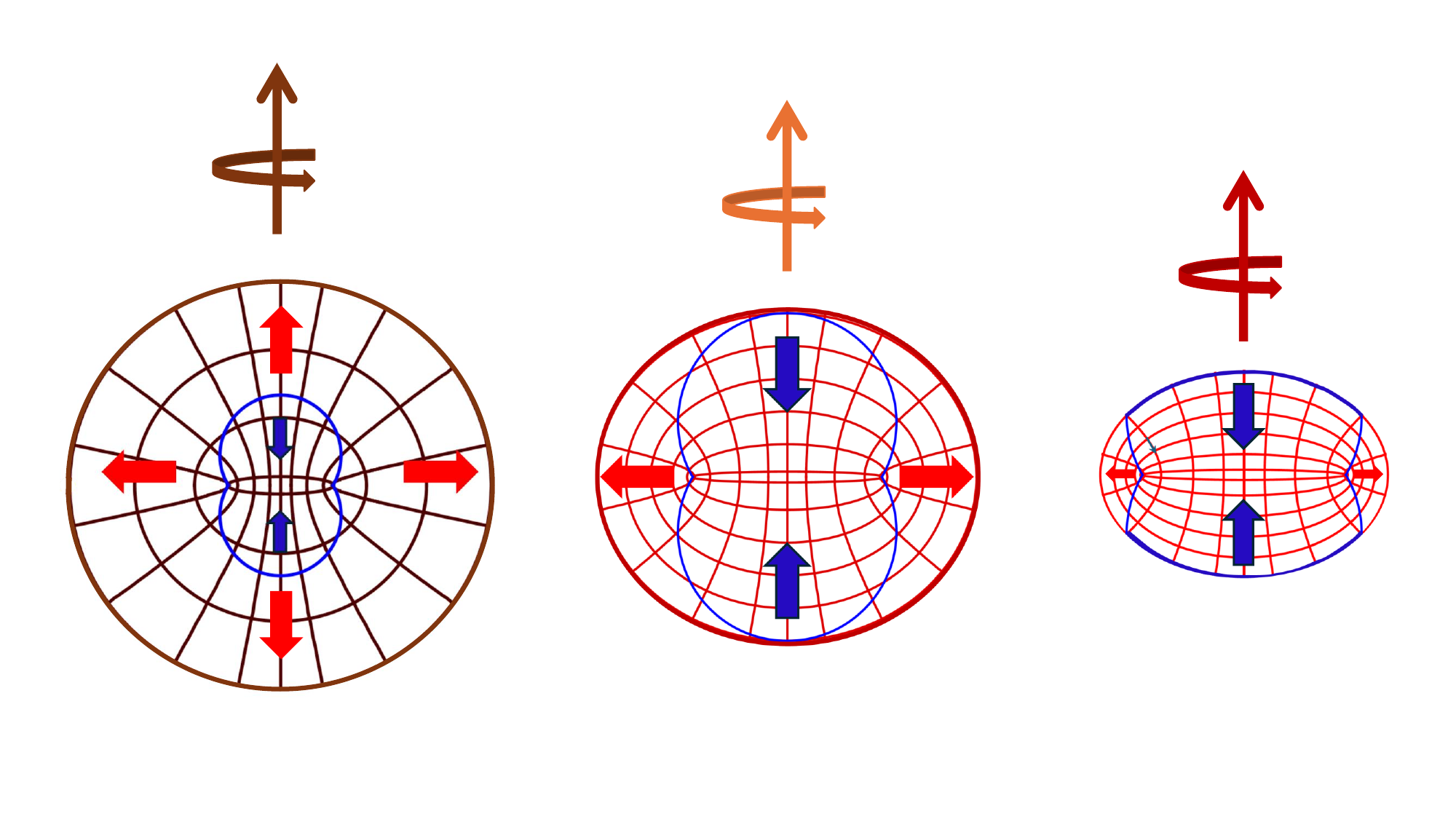}
\end{flushright}
\vspace{-.5in}
	\caption{Shown are cross-sections of fixed $\phi$ of the frozen star for the cases $a=0.3$ (left), $a=\sqrt{3}/2$ (middle) and $a=1$ (right).  The star is sourced by a fluid of electric-flux lines that either emanate from  or end at an ellipsoidal charge distribution in its core or, alternatively, emanate from the blue curve and end either at the outer surface or the center of the star. The flux is directed along the constant-$\theta$ hyperbolae. The flux inside the regions bounded by the blue curves is negative, as indicated by the blue, inward-pointing arrows, while the flux outside this region is positive, as indicated by the red, outward-pointing arrows. The outer-surface charge distribution has the same angular distribution as that in the core, but is oppositely charged. The flux lines are rigid and, consequently, the star is ultrastable.}
\label{fig}
\end{figure}

We will  next  discuss the charge  at the ends of the flux tubes, starting with  the one at the outer  surface of the frozen star. Let us  choose a narrow  Gaussian surface
whose sides are radially directed with one end just inside the outer surface  and the other end just outside the outer surface where $\;D_i=0\;$ for any $i$. Then using the Gauss'-law constraint~(\ref{sourceD}),  we find a  surface charge density of $\;\sigma(r_+)= -D_r\;$.
Hence, an outgoing flux leads to a negative surface charge density, while an ingoing flux leads to a positive charge density.
The total charge at the surface $q(r_+)$ is then obtained by integrating $D_r$,
\be
 q(r_+)\;=\; 2\pi \int_{-1}^{1} d\left[\cos\theta\right]\; D_r(r_+,\theta)\;=\;
 -\frac{\pi  \alpha '}{G} \;.
\label{tCharge}
\ee

In choosing the right-hand side of Eq.~(\ref{solD}) to be positive for small $a$,  we have implicitly chosen the signs of the charges to correspond to the case of a positive total charge at the star's center and, therefore, a negative total charge at its outer surface.  What is more interesting is the magnitude:  The total charge on the surface of a rotating star of any mass is precisely equal to that
of a static star of any mass \cite{fluxUstat}.

In the core, the calculation of the charge is more subtle because the geometry degenerates in the limit $\;r\to 0\;$. Therefore, we assume that the charge resides at  an infinitesimal but non-zero  radius $\;r=\delta\;$. Then choosing a radially directed, narrow Gaussian surface  with one end  at $\;r=\delta\;$  and the other close to  the outer surface of the star at $\;r=r_+\;$, we find that the surface densities are exactly equal in magnitude, $\;\sigma(\text{core})=-\sigma(r_+)\;$,  as previously argued. Hence,       $\;\sigma(\text{core})= +D_r\;$. Notice that the sign of the charge at any given point  is determined by the sign of $r^2-3 a^2 \cos^2\theta$.  Most of the flattened ellipsoid at $\;r=\delta\;$ is negatively charged, while the positive charge is pushed to a narrow ring for which  $\;\theta\to \pi/2\;$ with increasing $a$. Regardless, the total charge remains constant,
\be
q(\text{core})\;= \;\frac{\pi  \alpha '}{G}\;.
\label{tCharge2}
\ee

The charge distribution on the outer surface can be expanded in multipoles. For small $a$, the charge distribution is dominated by the monopole. The first correction is a negative quadrupole with a relative strength that scales as $a^2/R_+^2$.
This can be clearly seen in Fig.~1.

The emerging picture is compelling: When the star is at rest, the flux lines are straight, starting from a point-like charge at the star's center and ending at the outer surface. Then,  as the star begins to  rotate, the flux lines bend and are pushed away from the center to the surface where $\;r^2-3 a^2 \cos^2\theta=0\;$  due to the influence of a centrifugal force, as depicted in Fig.~1. Correspondingly, the positive, point-like charge spreads from the center to a ring of radius $a$ on the equator, while  the void near the axis of rotation is filled by  negative charge --- just enough  to   ensure that  the total charge remains the same. The ring of positive charge is at the same position as that of the would-be ring-singularity of the Kerr geometry. For small $a$, the resulting distribution is mostly quadrupolar. When the rotation is fast enough, the ring of positive charge is pushed further out,  whereas the region of negative charge  reaches the outer surface and continues  to spread  over it as the rotation approaches  its maximal value of $\;a=M\;$.

\section{Stability}

In this section, we discuss the stability of the rotating frozen star, beginning with its stability against the radial class of linear  perturbations. As expected, the rotating frozen star is shown to be ultrastable; meaning that all  linear perturbations of both the matter and metric vanish identically. So that, when in equilibrium,  a  rotating frozen star is just as stable as a  Kerr BH; neither can be excited, although their mass and/or angular momentum can  change. Then we discuss the possibility of ergoregion  instabilities --- those associated with negative-energy regions  inside
either the star or its external ergosphere --- and argue that any such instability could only develop over a time scale
that is much larger than the age of the Universe.

\subsection{Ultrastability against radial perturbations}

The property that ensures the ultrastability of the static frozen star,  $\;G^t_{\;\;t}=G^r_{\;\;r}\;$,  remains intact, and so we anticipate that the interior of the rotating frozen star is similarly ultrastable.

To see this explicitly, let  us start by linearly perturbing  $\;G_{\mu\nu}\to  G_{\mu\nu}+\delta G_{\mu\nu}\;$ and $\;T_{\mu\nu}\to  T_{\mu\nu}+\delta T_{\mu\nu}\;$. Then
\be
\frac{1}{8\pi G}\delta G_{\mu\nu}\;= \; \delta T_{\mu\nu}\;.
\ee

Let us next consider the various contributions to the right-hand side of the previous equation.
For $\;i$, $j$ = $\theta$, $\phi\;$ (but not $r$), it follows from Eq.~(\ref{Toi}) that
\be
\delta T_{0i}\;=\; -\frac{1}{2\pi\alpha'}\delta {\cP}_i
\label{dToi}
\ee
and from Eq.~(\ref{Tij}) that
\be
(2\pi \alpha')^2 \cH \; \delta T_{ij}\;= \;-\delta (D_i D_j) + \delta ( {\cP}_i {\cP}_j) + (2\pi\alpha'\cH)^2 \; T_{ij}\;\delta \left(\frac{1}{\cH}\right)  \;=\; 0\;,
\label{dTij}
\ee
with the right-most equality due to the vanishing of the background values of the  $D_i$'s, ${\cP}_i$'s and $T_{ij}$'s.  Notice that this last result indicates that fluctuations
in the transverse pressure vanish, $\;\delta p_{\perp}=0\;$.

As for  $\delta T_{rj}$ and $\delta T_{r0}$, let us start with
\be
(2\pi \alpha')^2\delta \left( \cH T_{rj}\right) \;=\; - \delta (D_r D_j) + \delta ( {\cP}_r {\cP}_j)\;.
\ee
Then, since the background values of  the $T_{rj}$'s, $D_j$'s and any of the $\cP$'s vanish,
\be
\delta T_{rj} \;=\; - \frac{1}{2\pi\alpha'}\delta D_j\;,
\label{dTrj}
\ee
where we have also used that, on the solution, $\;2\pi \alpha' \cH= \sqrt{D^rD_r}=D_r\;$,  with the latter equality because of our choice of convention.
Next,
\be
\delta T_{r0}\;= \;-\frac{1}{2\pi\alpha'}\delta \cP_r \;= \;\delta (D\times B)_r \;=\;0\;,
\label{dTor}
\ee
where the final equality comes about because, on the solution,  $D_r$ is the only non-vanishing component
of the  electric displacement and  all components of the magnetic induction vanish.

Moving on to the diagonal components of the $t,r$ sector and again using that, on the solution,
$D_r$ is the only non-vanishing component of the displacement and
$\;2\pi \alpha' \cH = \sqrt{D^rD_r}=D_r  \;$,
we have
\be
\delta T^0_{\ 0} \;= \;\delta \cH
\;=\; \frac{1}{2\pi\alpha'}\delta D_r \label{fT0}
\ee
and
\bea
\delta T^r_{\ r} &\;=\;& -\frac{1}{2\pi\alpha'}\left[2\delta D_r - \frac{1}{2}\frac{\delta(D^aD_a)}{\sqrt{D^aD_a}}\right] \cr &\;=\;& -\frac{1}{2\pi\alpha'}\left[2\delta D_r -
\delta D_r\right]
\;=\;-\frac{1}{2\pi\alpha'}\delta D_r\;,  \label{fTr}
\eea
where, in the second line, the diagonality of  the on-solution
metric has also been used and  the vanishing  ${\cal P}^2$
contribution has now been discarded from the outset.
In other words,
\be
\delta T^0_{\ 0} + \delta T^r_{\ r} \;=\; 0\;,
\ee
meaning that  the equation of state is preserved for linear perturbations.

We now pivot to the left-hand side of the linearized Einstein equations; that is,
the calculation of $\delta G_{\mu\nu}$.
This can be done by perturbing the metric by four functions of $t$ and $r$,
as is the standard practice for even-parity perturbations \cite{thorne}.
Denoting the four functions by $H_0$, $H_1$, $H_2$ and $K$, we can then write
\be
ds^2\;=\; -\varepsilon^2(1+ H_0)dt^2 + 2 \partial_t H_1 dt dr + \frac{1}{\varepsilon^2}(1+ H_2) dr^2 + (1+ K)\left[ \Sigma d\theta^2 + \frac{R^2}{\Sigma} \sin^2\theta d\phi^2\right].
\ee
Let us  parameterize the time dependence of all perturbations as $e^{i \omega t}$ and
further assume that $\;\omega\ne 0\;$. The stability for the case that the perturbations do not depend on $t$, but only on $r$, is simpler because $K$ can be set to zero by an $r$-dependent coordinate transformation.

We proceed by expanding $G_{\mu\nu}$ to linear order in the perturbations and to leading order in $\varepsilon^2$. The calculation is quite involved and  {\em Mathematica}
has been used to perform the expansion. The results for the diagonal components are as follows (where a subscript
of $0$ on a density  denotes a background quantity):
\be
G^0_{\ 0}\;= \;8\pi G\rho_0~ K\;,
\label{pertTT}
\ee
\be
\delta G^r_{\ r}\;=\;- 8\pi G(p_r)_0~ K +\frac{2r}{R^2}\partial_t^2 H_1-\frac{1}{\varepsilon ^2}{\partial_t^2}K\;,
\label{pertRR}
\ee
\be
\delta G^{\theta }_{\ \theta }\;=\; \delta  G^{\phi }_{\ \phi }\;=\; -\frac{1}{2}\frac{1}{\varepsilon ^2}\left(\partial^2_t {H_2} +\partial^2_t K \right)\;.
\label{pertIJ}
\ee

We also find that, for $\;i=\theta, ~ \phi\;$,
\be
\delta G_{0 i} \;=\;\delta G_{r i}\;=\;0 \;
\label{pertTI}
\ee
and  that
\be
\delta G_{0r}\;=\; 8\pi G\rho_0 \partial_t H_1 +\frac{1}{R^2}\partial_t\left(r H_2 -r K + R^2 \partial_r K\right)\;.
\label{pertTR}
\ee

To summarize, we obtain, along with $\;\delta D_i=0\;$ and $\;\delta{\cal P}_i=\delta{\cal P}_r=0\;$, the following set of linearized perturbation equations:
\bea
&& \rho_0 K \;=\; \frac{1}{2\pi \alpha'} \delta D_r\;, \\
&&\rho_0 K +\frac{1}{8\pi G}\left[\frac{2r}{R^2}\partial_t^2 H_1-\frac{1}{\varepsilon ^2}{\partial_t^2}K\right]\; = \;
-\frac{1}{2\pi\alpha'}
\delta D_r\;, \\
&& \rho_0 \partial_t H_1 +\frac{1}{8\pi G}\left[\frac{1}{R^2}\partial_t\left(r H_2 -r K + R^2 \partial_r K\right) \right] \;= \;0\;, \\ &&
-\frac{1}{2}\frac{1}{\varepsilon ^2}\left(\partial^2_t {H_2} +\partial^2_t K \right)\;=\;0\;.
\eea
 From the second equation, it follows that $K$ vanishes, otherwise the third term on the left blows up with nothing to compensate for it.  Then, from the first and fourth equation, it must be that  $\delta D_r$  and
$H_2$ respectively vanish. From the vanishing of $K$ and $H_2$, it follows
from the third equation that $\;H_1=0\;$. Since $\delta D_r$ vanishes, so must
$\delta \rho$ and $\delta p_r$ ({\em cf}, Eqs.~(\ref{fT0},\ref{fTr})), as did $\delta p_{\perp}$.

Our conclusion is that all matter perturbations and metric perturbations vanish identically, to leading order in $\varepsilon^2$, except for $H_0$, which is left undetermined.~\footnote{The induction fluctuations, $\delta B_i$ and $\delta B_r$, can only enter the formalism via the momenta fluctuations, all of which vanish.}  The reason that $H_0$ evades determination can be traced back to the possibility of eliminating it by a transformation of the time coordinate, which does not change the other components of the metric to leading order in $\varepsilon^2$.  A similar phenomenon was encountered in the analysis of radial stability in the static case. To establish the vanishing of $H_0$,  one needs to allow deviations from  $\;p_r=-\rho\;$ and then take the limit $\;p_r \to -\rho\;$
{\em a posteriori}.

\subsection{Absence of ergoregion instabilities}

As previously observed in Section~3, the rotating frozen star contains regions of
negative energy, or ergoregions, in its interior. These ergoregions are different than the ergoregion of the exterior Kerr geometry, which is also shared by the frozen star.

The exterior ergoregion of horizonless, rotating, ultracompact objects  leads to  well-known  instabilities \cite{MoreCardoso}. The physical origin of such an instability is that a wave or particle that is scattered by the ultracompact object can extract rotational energy from it.  A region of negative energy can then lead to superradiant scattering, meaning that a scattered wave can end up with  a larger amplitude than that of the original  incident wave (see, for example, \cite{BekSchif}).  This effect will lead to instabilities when the scattering process happens repeatedly; enough times so that the amplitude of the multi-scattered  wave can grow exponentially large.

As a horizonless, rotating, ultracompact object, a  frozen star might well be suffering from these types of  instabilities. However, as argued next, the timescale of any such instability would be extremely large for an astrophysically relevant frozen star, and so  the effects of the instability would be irrelevant for all practical purposes.

For an exterior instability to be relevant, the wave has to traverse it many times. However, as explained below, the absorption coefficient of a macroscopic frozen star is effectively unity, up to a correction of order $\varepsilon^2$, so that an infalling wave effectively crosses the external ergosphere only once.

Let us recall from \cite{trajectory} that the light-crossing time for a solar-sized frozen star was estimated to be $10^{32}$~s, which is longer than the lifetime of the Universe! This  estimate  can be attributed to waves or particles having interior velocities on the  order of the extremely small parameter $\varepsilon^2$, whose magnitude is of the order of the ratio of the Planck length to the (would be) horizon length scale, $\;\varepsilon^2\sim \frac{l_P}{
r_+}\;$. This is because waves are  restricted  to travelling along  radial, null trajectories, $\;\frac{dr}{dt}=\varepsilon^2\;$. Since the requisite trip time has to be much, much  larger than the light-crossing time, this all but rules out the possibility of this type of instability for any frozen star whose size is of astrophysical relevance.

The argument is valid for both the exterior and interior ergoregions of the rotating frozen star.
To understand why the interior ergoregion is not an issue,  consider that  the scattering surface in this case can either be at the outer layer or in the central core of the star. Furthermore, such an  instability could only happen after a given wave has transversed the interior of the  star many times. Then, since the crossing time is extremely large, the huge time scale of the instability once again renders it irrelevant for astrophysical objects.

\section{Smooth connection to the exterior Kerr metric and regularization at the core}

\subsection{Matching the frozen star and Kerr geometries at the outer surface}

To match continuously the interior geometry of the frozen star to the near-horizon, external
Kerr geometry, what is required is   a narrow transitional layer  of width $\;\lambda\ll r_+\;$ (but $\;\lambda\gg \varepsilon^2 r_+$), as this allows for the interior metric and its first two derivatives --- what is needed to describe the Einstein tensor --- to be smoothly connected to the exterior. Such a procedure has already been carried out for the
spherically symmetric, static star, as documented in \cite{popstar}.

In practice, we need to find four interpolating polynomials
$P_{tt}$, $P_{rr}$, $P_{t\phi}$ and  $P_{\phi\phi}$
for each of the four  metric components $g_{tt}$, $g_{rr}$, $g_{t\phi}$ and  $g_{\phi\phi}$. Note that $g_{\theta\theta}$ is already continuous and does not require further consideration. The polynomials need to be of degree 5 so as to ensure the continuity of the metric and its first two derivatives at both ends of the transition layer.

It is useful to recall Eq.~(\ref{offdiagonal1}) for the frozen star metric
in its Boyer--Lindquist-like form, as this will be the choice used in the matching procedure. It can be rewritten as
\bea
{ds_{FS}^2}&=&\left(\frac{a^2\sin^2\theta}{\Sigma}-\varepsilon^2\right) dt^2 -
\left(\frac{4\, M\, r\, a}{\Sigma}+  2 a\,\frac{\Delta}{\Sigma} \right) \sin^2\theta dt d\phi ~ +\frac{1}{\varepsilon^2}dr^2 \cr &+&\Sigma d\theta^2
+ \frac{R^4}{\Sigma}\sin^2\theta d\phi^2\;.
\eea
Let us, for convenience, start the transitional  layer at $\;r=r_+\;$,
where $\;\Delta=R^2-2Mr=0\;$ or $\;R^2=2 M r\;$.  We can use this last result to simplify the metric components and their first and second derivatives at $r_+$.

The idea is then to smoothly connect the frozen star metric over a distance $\;\lambda\ll r_+\;$ to the exterior Kerr metric at $\;r=r_++\lambda\;$.~\footnote{In this regard and also including corrections of order $\varepsilon^2$, the interior geometry formally stops  at $r_+$
and the  exterior geometry formally begins  at  $r_+  + {\lambda}$. Both of these differ perturbatively from $r_++\varepsilon^2$, which (along with $r_+$) is sometimes used as an
estimate of the star's radius.} In Boyer--Lindquist coordinates, as is appropriate for the matching, the latter  is
\bea
ds^2_{Kerr}&=&-\left(1-\frac{2 M r}{\Sigma}\right)dt^2 - \frac{4 \, M\, r\, a }{\Sigma}\sin^2\theta dt d\phi
+\frac{\Sigma}{\Delta}\, dr^2 \cr &+& \Sigma d\theta^2+ \left( R^2 + \frac{ 2 \, M\, r\, a^2 \sin^2\theta}{\Sigma} \right)\sin^2\theta d\phi^2\;.
\eea
Using $\;\Sigma=r^2+a^2\cos\theta^2\;$ and $\;\Delta=R^2-2 \,M\, r \;$, we can simplify the first
and last components of the above metric. Respectively,
\be
1- \frac{2 M r}{\Sigma}\;=\; \frac{\Delta}{\Sigma}-\frac{a^2\sin^2\theta}{\Sigma}\;,
\ee
and
\be
R^2 +  \frac{2 \, M\, r\, a^2 \sin^2\theta}{\Sigma}\;=\;\frac{R^4}{\Sigma}-{a^2\sin^2\theta}\,\frac{\Delta}{\Sigma}\;.
\ee

To summarize,
\bea
 {ds_{FS}^2}&=&\left(\frac{a^2\sin^2\theta}{\Sigma}-\varepsilon^2\right) dt^2 -
\left(\frac{4 \, M\, r\, a }{\Sigma}+2 a\, \frac{\Delta}{\Sigma}\right)\sin^2\theta dt d\phi ~ +\frac{1}{\varepsilon^2} dr^2 \cr &+&   \Sigma d\theta^2
+ \frac{R^4}{\Sigma}\sin^2\theta d\phi^2\;, \\
ds^2_{Kerr}&=&\left(\frac{a^2\sin^2\theta}{\Sigma}-\frac{\Delta}{\Sigma}\right) dt^2 - \frac{4 \, M\, r\, a }{\Sigma}\sin^2\theta dt d\phi
+\frac{\Sigma}{\Delta}\, dr^2 \cr &+& \Sigma d\theta^2+ \left( \frac{R^4}{\Sigma}-{a^2\sin^2\theta}\,\frac{\Delta}{\Sigma} \right)\sin^2\theta d\phi^2\;.
\eea

The conclusion is that, to perform the smoothing, we need  the polynomial  expansion of $\dfrac{\Delta}{\Sigma}$ (and  $\dfrac{\Sigma}{\Delta}$) to second order in $r$-derivatives. We denote this polynomial by $Q_{ds}(r,\theta)$ and use it to smoothly connect $g_{tt}$, $g_{rr}$, $g_{t\phi}$ and $g_{\phi\phi}$, with $g_{\theta\theta}$  smooth by construction. Then, wherever there is a difference proportional to $\dfrac{\Delta}{\Sigma}$  between the frozen star  and  Kerr  metrics, we replace it by  a polynomial of the form $\varepsilon^2+A(r-r_+)^3+B(r-r_+)^4 +C (r-r_+)^5 $, with $A,~ B,~ C$ determined by the metric component and its first two derivatives at $\;r=r_++\lambda\;$.

For example, the polynomial $P_{rr}$ for $g_{rr}$ has to satisfy the following six relations,
\bea
&& \left(P_{rr},~P'_{rr},~P''_{rr}\right)_{|(r_+ )}\;=\;\left(\frac{1}{\varepsilon^2},0,0 \right)\;, \\
&&  \left(P_{rr},~P'_{rr},~P''_{rr}\right)_{|(r_+ +\lambda )}\;=\;\left(\frac{\Sigma}{\Delta},\left(\frac{\Sigma}{\Delta }\right)', \left(\frac{\Sigma}{\Delta}\right)'' \right)_{|(r_+ +\lambda)}\;.
\eea
The end result of this process are four polynomials of degree 5 which smoothly interpolate   the metric and Einstein tensor from the interior of the frozen star to the exterior Kerr geometry, while maintaining the exact angular dependence.

The resulting components of the Einstein tensor, and therefore the stress tensor,
scale in a similar way to their counterparts in the static case. The transverse pressures $p_\theta$, $p_\phi$ are equal and vanish at both ends of the transitional layer, while scaling  in the middle of the layer as $1/\lambda$. The energy density $\rho$ and magnitude of the radial pressure $|p_r|$ decrease from their common frozen star value at $\;r=r_+\;$ to zero at $\;r=r_+ +\lambda\;$. Both are independent of $\lambda$ at leading order.

The full analysis for the matching procedure is presented in an appendix.
The calculations are complicated by the fact that $\Delta$ is a small quantity near the outer surface at $r_+$, but its first and second derivatives are not, so that the orders of the perturbative expansions  mix for the first and second derivatives of the metric components.

\subsection{Regularizing the core}

As can be seen in Section~3, the curvature and all non-vanishing components of the Einstein tensor remain regular throughout the interior, except for a mild singularity near the center of the star, similarly to  the spherically symmetric, static case.  In spite of the singular behavior, the amounts of mass and angular momentum in the core region  are perturbatively small, as revealed by direct integration of the energy and angular-momentum densities. (An    explicit expression for the latter can be found in \cite{notstevekerr}.) In fact, the situation is even better for the rotating case because, for a non-vanishing $a$, the singularity  is restricted to the equator. Still, it is useful to confirm that a suitable regularization procedure can indeed be enforced.

It turns out that it is enough to regularize the metric by changing the angular function $\;\Sigma=r^2+a^2\cos^2\theta\;$  in a small core region near $\;r=0\;$. For example, $\Sigma$ is modified for $\;r \le \eta\;$ with $\;\eta\ll r_+\;$ and $\;{\eta}\gg\varepsilon^2 {r_+}\;$, while $\Sigma$ is unchanged for $\;r\ge \eta\;$. Then, in this core region, we redefine  $\;\Sigma \to\Sigma+f(r)\;$ such that $\;f(0)=const.\;$ and $f\;(r=\eta)=0\;$,       $\;f' (r=\eta)=0\;$ and $\;f^{''}(r=\eta)=0\;$, so that the metric matches smoothly to the bulk metric of the frozen star. A simple example for such a regulator is
\be
f(r)=
\begin{cases}
  f(r)=\frac{(\eta -r)^3}{\eta } & r\le \eta\;,  \cr 0 & r\ge \eta \;.
\end{cases}
\ee

The resulting Einstein tensor $G^{a}_{\ b}$, to leading order in $\varepsilon^2$, is almost identical to the unregularized tensor, with the only order-unity components (after diagonalization) being $G^t_{\ t}$ and $\;G^r_{\ r}=-G^t_{\;t}\;$. Via Einstein's equations, this equality maintains the equation of state $\;p_r=-\rho\;$, but now with
\be
8\pi G \rho \;= \;\frac{\left(a^2+r^2+f(r)\right) \left(r^2-3 a^2 \cos ^2\theta +f(r)\right)}{\left(a^2 \cos^2\theta+r^2+f(r)\right)^3} \;.
\ee
The only difference between the regularized and unregularized cases is the appearances of $f(r)$ in the former. Notice that derivatives of $f(r)$ do not appear.
On the equator, the previously singular energy density  on the ring at $\;r=0\;$ and $\;\cos\theta=0\;$ now scales as
\be
\rho\;\sim\;\frac{a^2+f(0)}{f(0)^2}\;=\;\frac{a^2+\eta^2}{\eta^4}\;\sim\; \frac{a^2}{\eta^4}\;.
\ee
In the limit $\;a\to 0\;$, the energy density scales as $\frac{1}{\eta^2}$, as
it did in the static case.

As for the flux, since $\;D_r\propto \rho\;$, the result of the regularization is to modify what looked like a singular ring of charge to an ellipsoid of charge with  regions of positive and negative charge distributions, as discussed in Section 3.2.

From \cite{notstevekerr}, we know that the mass in the regularized core region scales as $\;M(r\le\eta)= \tfrac{1}{2} \eta\;$, so that $\;\frac{M(r\le\eta)}{M}=\frac{\eta}{r_+}\ll 1\;$, and  similarly for the ratio of the angular momentum in the core to the total angular momentum, $\;\frac{J(r\le\eta)}{J}=\frac{\eta}{r_+}\ll 1\;$.

\section{Discussion}

We have shown how the Born--Infeld framework, which was introduced in \cite{fluxUstat}, can be extended from the case of the  spherically symmetric, static frozen star to its  axially symmetric, rotating counterpart, which was first proposed in
\cite{notstevekerr}. The importance of  having an explicit form of matter  Lagrangian cannot be overstated. It allows  us to study frozen star geometries that have no such symmetries and
consider the evolution of these solutions and their stability in fully dynamical scenarios. This will be particularly useful for modeling the early phases of BH mergers, which are expected to be observationally  relevant for the purpose of  discriminating  between frozen stars, Kerr BHs and also between frozen stars and other BH mimickers \cite{collision,ridethewave}.

An important next step in the just-discussed program  would be to extend this
framework to the case of the  defrosted star \cite{fluctuations}, for which $\;\rho+p\;$ and $p_{\perp}$ are no longer vanishing but perturbatively small, yet much larger than $\varepsilon^2$.  Deviating  from the $\;\rho+p=0\;$ equation of state is necessary if we are to study the  star's ringdown modes and resulting gravitational-wave emissions because of the ultrastability of the solution for the case $\;\rho+p=0\;$ .
This was made clear in \cite{fluctuations}, where it was shown how the defrosted star can support excitations whereas the fully  frozen star cannot.
To extend the framework in this way  will, however, entail another level of complexity. In addition to a weak, radially directed electric field, a  radially directed magnetic field will also be required.  Such a magnetic field can be sourced by a magnetic monopole,
meaning that the complete solution is sourced by a dyon. Work along this line is already in progress.

\section*{Acknowledgments}
We thank  Yoav Zigdon for discussions, Suvendu Giri for pointing to us the similarity of the frozen star model to the clouds of string model and to the Born-Infeld BIons and Frans Pretorius for insisting that we find the source matter Lagrangian of the frozen star. We extend our special thanks to Piljin Yi, for helping us understand his work and its implications to the frozen star model. The research is supported by the German Research Foundation through a German-Israeli Project Cooperation (DIP) grant ``Holography and the Swampland'' and by VATAT (Israel planning and budgeting committee) grant for supporting theoretical high energy physics. The research of AJMM received support from an NRF Evaluation and Rating
Grant 119411 and a Rhodes  Discretionary Grant SD07/2022.
AJMM thanks Ben Gurion University for their hospitality during past visits.

\appendix
\section{Smoothing of the outer transitional layer}

Here, we provide more details about the smoothing procedure that made use of an outer transitional layer to connect the frozen star interior to the Kerr exterior.

In the outer transition layer, we use the Kerr metric with $\frac{\Delta}{\Sigma}$ replaced by the degree-5 polynomial $Q_{ds}(r,\theta)$. The resulting line element is given by
\bea
ds^2_{TL}&=&\left(\frac{a^2\sin^2\theta}{\Sigma}-Q_{ds}(r,\theta)\right) dt^2 + 2 a
\sin ^2\theta \left(Q_{ds}(r,\theta )-\frac{R^2}{\Sigma}\right)
dt d\phi \nonumber \\
&+&\frac{1}{Q_{ds}(r,\theta)}\, dr^2  + \Sigma d\theta^2+
\left( \frac{R^4}{\Sigma}-{a^2\sin^2\theta}\,Q_{ds}(r,\theta) \right)\sin^2\theta d\phi^2\;. \nonumber
\label{TLds}
\eea

The interpolating polynomial $Q_{ds}$  and it first and second radial derivatives  at the inner boundary of the transitional layer, which is located  at $\;r=r_+\;$, have to
agree with the frozen star geometry, leading to  the three conditions \\ $\;\left(Q_{ds}(r_+,\theta)= \varepsilon^2\;,\;Q_{ds}'(r_+,\theta)=0\;,\; Q_{ds}''(r_+,\theta)=0\right)\;$. Meanwhile, at the outer boundary of the layer,  where
$\;r=r_++\lambda\;$,
the polynomial and its derivatives must satisfy conditions that are based
on the external Kerr solution,  $$\left(Q_{ds}(r_++\lambda,\theta)= \frac{\Delta}{\Sigma}_{|r=r_++\lambda},Q_{ds}'(r_++\lambda,\theta)= \left(\frac{\Delta}{\Sigma}\right)'_{|r=r_++\lambda}, Q_{ds}''(r_++\lambda,\theta)= \left(\frac{\Delta}{\Sigma}\right)''_{|r=r_++\lambda}\right)\;.$$

The resulting polynomial can be expressed in terms of $\;x=r-r_+$\;, $\;\Sigma_{+\lambda}= (r_++\lambda )^2+ a^2 \cos ^2\theta\;$ and  $\;Q_{ds}(x,\theta)=\varepsilon^2+ A~\dfrac{x^3}{\lambda ^3}+B~ \dfrac{x^4}{\lambda ^4}+C~ \dfrac{x^5}{\lambda ^5}\;$, where the coefficients $A$, $B$, $C$
work out to be~\footnote{Some of these can be further simplified by way of $\;\Delta(r_+)=r_+^2- 2 M r_+ + a^2=0\;$.}
\bea
\cr
A &=&
10 \left(\frac{a^2+(r_++\lambda ) (r_++\lambda -2 M)}{\Sigma_{+\lambda} ^2}-\varepsilon ^2\right)
\cr &-&
8 \lambda\frac{  a^2 \cos ^2\theta (r_++\lambda -M)+(r_++\lambda ) \left(M (r_++\lambda )-a^2\right)}{\Sigma_{+\lambda} ^2}
\cr &+&
\lambda ^2 \frac{ a^4 \cos ^4\theta-a^2 \cos ^2\theta \left(a^2+3 (r_++\lambda ) (r_++\lambda -2 M)\right)+(r_++\lambda )^2 \left(3 a^2-2 M (r_++\lambda)\right)} {\Sigma_{+\lambda}^3}\;,
\cr
B&=& \frac{16 \lambda  (M-r_+)-3 \lambda ^2}{\Sigma_{+\lambda}}
\cr &-&
2 \lambda \frac{  a^2 (7 r_++6 \lambda )+\lambda (r_++\lambda ) \left(  (9 r_+-8 M)+7 r_+ (r_+-2 M)+2 \lambda ^2\right)}{\Sigma_{+\lambda}^2}
\cr &-&
8 \lambda ^2\frac{ (r_++\lambda )^2 \left(a^2+(r_++\lambda ) (r_++\lambda -2 M)\right)}{\Sigma_{+\lambda}^3}+15 \varepsilon ^2 \;,
\\
C&=&\frac{6 \lambda  (r_+-M)+\lambda ^2}{\Sigma_{+\lambda}}+
\cr &+&
\lambda  \frac{ a^2 (6 r_++5 \lambda )+\lambda (r_++\lambda ) \left(  (7 r_+-6 M)+6 r_+ (r_+-2 M)+\lambda^2\right)}{\Sigma_{+\lambda}^2}
\cr &+&
4 \lambda ^2\frac{ (r_++\lambda )^2 \left(a^2+(r_++\lambda ) (r_++\lambda -2 M)\right)}{\Sigma_{+\lambda}^3}-6 \varepsilon ^2\;.  \nonumber
\eea

We have verified explicitly that the resulting components of the Einstein tensor --- equivalently, the stress tensor --- at $\;r=r_+\;$ are indeed equal to their frozen star counterparts and that they vanish at $\;r=r_++\lambda\;$ as required.

Due to the complicated expression for the interpolating polynomial $Q_{ds}(r,\theta)$, it is not practical nor particularly useful to present the expressions for the Einstein tensor components as a function of $r$ and $\theta$ throughout  the transitional region.  We do, however, present their values at the midpoint of the  layer, $\;r = r_++\frac{\lambda}{2}\;$, to leading order in $\;\frac{\lambda}{r_+}\;$.~\footnote{The extremal limit $\;r_+\to a\;$ has to be treated with care.} To arrive at these values, we calculated the Einstein tensor resulting from the metric in Eq.~(\ref{TLds}), then substituted the leading-order values of the polynomial and its first and second derivatives with respect to $r$ and $\theta$ at the midpoint. We then expanded the result in $\lambda/r_+$ and noted the leading-order value.
The resulting set of non-vanishing energy-momentum components at the midpoint are
\bea
G_{tt} &=&\frac{3}{4 \lambda r_+} \frac{a^2 \left(r_+^2-a^2\right) \sin ^2\theta}{   \Sigma_+^2}\;,
\\
G_{t\phi} &=&-\frac{3}{4 \lambda r_+}\frac{ a \left(r_+^4-a^4\right) \sin^2\theta}{  \Sigma_+^2}\;,
\\
G_{\phi\phi} &=&\frac{3}{4 \lambda r_+} \frac{\left(a^2+r_+^2\right) \left(r_+^4-a^4\right) \sin ^2\theta}{   \Sigma_+^2}\;,
\\
G_{rr} &=&\frac{14 }{11 \lambda} \frac{r_+}{ \Sigma_+}\;,
\\
G_{r\theta} &=&-\frac{34}{11}\frac{ a^2 r_+ \sin (2 \theta )}{ \Sigma_+^2}\;,
\\
G_{\theta\theta} &=&\frac{3}{4 \lambda r_+}\left(r_+^2-a^2\right)\;.
\eea

After diagonalizing the $\theta,r$ sector, one  can   observe that, to leading order in $\lambda/r_+$,
\be
G^\theta_{\ \theta}\;=\;G^\phi_{\ \phi}\;=\; \frac{3}{4 \lambda r_+}\frac{ \left(r_+^2-a^2\right)}{ \Sigma_+}
\ee
and
\be
G^r_{\ r}\;=\;\frac{7 \left(r_+^2-a^2\right)}{16 \Sigma_+^2}\;,
\ee
where $\;\Sigma_+=\Sigma(r_+)\,$.

Furthermore, we observe that $\;G_{tt} = -\Omega_H G_{t\phi}= \Omega_H^2 G_{\phi\phi}\;$.
Diagonalizing the $t,\phi$ sector therefore requires one to expand the various quantities to a higher order in $\lambda$, which once again makes for a  completely opaque result. However, it is clear that, after diagonalization, $\;\rho=-G^t_{\ t}\;$
(in $\;8\pi G=1\;$ units) is positive, independent of $\lambda$ at leading order and not equal to $-p_r$ except at the ends of the layer.

\end{document}